\documentclass[proof]{WileyASNA-v1}

\articletype{Article Type}%

\received{XX September 2021}
\revised{XX XX 2021}
\accepted{XX XX 2021}

\begin{document}

\title{Stellar rotation and its connection to the evolution of hydrogen-dominated atmospheres of exoplanets}

\author[1,2]{Daria Kubyshkina*}



\authormark{Daria Kubyshkina}

\address[1]{\orgdiv{School of Physics}, \orgname{Trinity College Dublin, the University of Dublin}, \orgaddress{\state{Dublin}, \country{Ireland}}}

\address[2]{\orgdiv{Space Research Institute}, \orgname{ Austrian Academy of Sciences}, \orgaddress{\state{Graz}, \country{Austria}}}


\corres{* \email{kubyshkd@tcd.ie}}

\presentaddress{College Green, Dublin-2, Ireland}

\abstract{The population of known low- to intermediate-mass exoplanets shows a large spread in densities, which is believed to be due to the diversity of planetary atmospheres and thus controlled by planetary atmospheric mass loss. One of the main drivers of long-term atmospheric escape is the absorption of high-energy XUV radiation from the host star. For main sequence solar-like stars, rotation and XUV radiation are closely connected, with faster rotating stars being XUV brighter and with both rotation and XUV decreasing with time. This evolution, however, does not follow a unique path, as stars born with the same mass and metallicity can have widely different initial rotation rates. This non-uniqueness holds up to about 1 Gyr, while atmospheric escape from exoplanets is strongest. The atmospheric mass loss through this period is often deciding the future of the planet and its position in the observed population. Therefore, the diversity of possible stellar histories can be an uncertain factor affecting the predictions of population studies. Here, I explore its relevance for different planets and different host stars.}


\jnlcitation{\cname{\author{Kubyshkina D.} (\cyear{2021}), \ctitle{Stellar rotation and its connection to the evolution of hydrogen-dominated atmospheres of exoplanets}, \cjournal{Astronomische Nachrichten}, \cvol{??}.}}


\maketitle


\section{Introduction}\label{sec::intro}

Numerous observations of exoplanets in the past decade have revealed the large groups of planets in parameter ranges not presented in the Solar system, including the close-in giants and intermediate size planets \cite[][]{mullally2015,fulton2017,vaneylen2018}. Among them, the population of planets with masses between that of Earth and Neptune shows a large spread in densities and a dearth of planets with radii of about 1.9~$R_{\oplus}$, referred to as an evaporation valley or a radius gap. The main shaping factor of this population is believed to be the atmospheric mass loss \citep[e.g.,][]{owen_wu2017,jin2014,jin_mordasini2018,sandoval2021}. This implies, that all planets in the population were born with substantial hydrogen-dominated atmospheres (explaining well the large radii of sub-Neptune counterparts), but some have lost it through time and became dense and compact super-Earths.

To date, however, the main driving mechanism of this atmospheric mass loss is debated.
The two leading approaches are the atmospheric evaporation due to high-energy stellar irradiation \citep[in particular, energy-limited approximation,][]{watson1981,owen_wu2013,owen_wu2017,jin2014} and so-called ``core-powered'' mass loss driven by the cooling of planetary core \citep{ginzburg2018,gupta_schlichting2019,gupta_schlichting2020}. Thought the imprints of atmospheric escape are found in the observed population \citep{david2021,sandoval2021}, and the two potential mechanisms are expected to produce different signatures \citep{gupta_schlichting2020}, the present day accuracy of observations does not allow distinguishing them \citep{rogers2021}.

Alternative theories explain the radius gap as a consequence of planetary formation processes, as the preceding protoplanetary nebula lifetime and atmospheric accretion \citep{stoekl2015}, core composition \citep{schiller2018,morbidelly2020,venturini2020,lee_connors2021}, planetary migration \citep{jin_mordasini2018,mordasini2020} and atmospheric erosion by impacts \citep{bonomo2019,wyatt2020}.
Even if not considering the radius gap as a direct product of planetary formation processes, they are responsible for the initial parameters of planets after the protoplanetary disk dispersal, when the atmospheric mass loss begins \citep{rogers_owen2021}.

In this study, I focus on the effect of atmospheric escape induced by high-energy stellar radiation (X-ray + extreme ultraviolet, XUV), which is closely connected to evolution of the host star.
It is known that rotation of a star and its activity (XUV) are bound \citep[e.g.,][]{mamajek2008,wright2011,jackson2012,magaudda2020}. Both decay with time, but this decay does not follow a unique path at ages before $\sim$1~Gyr \citep{tu2015,matt2015,johnstone2021}. This leads to a wide spread in stellar XUV luminosities, and thus to a very different amount of stellar XUV radiation received by companion planets through this period. Integrated over time, it can be orders of magnitude different for planets at similar orbits around stars of similar masses but evolved differently in terms of XUV.

This issue is further complicated by that for stars older than $\sim$1~Gyr it is not possible to know the past rotation (hence, XUV) solely from present-day stellar parameters.
\cite{kubyshkina2019} suggest the method allowing to resolve it for individual stars hosting sub-Neptune-like planets, which was further improved by Bonfanti et al. (2021, under review). Here, however, I focus on a more general question on how the uncertainty in the amount of XUV radiation received by planets throughout the first Gyr of their lives can affect predictions of population studies and, therefore, our understanding of planetary evolution and demographics.

I present the atmospheric mass loss rates predicted by hydrodynamic modeling for a wide range of planets in Section~\ref{sec::escape} and discuss the role of particular driving mechanisms through evolution in Section~\ref{sec::escape_evol}. In Section~\ref{sec::stellar_rot}, I describe the setup of stellar evolution models \citep{johnstone2021}, and discuss the relevance of uncertain stellar history for planetary population in Section~\ref{sec::population}. I outline the conclusions in Section~\ref{sec::conclusions}.

\section{Atmospheric mass loss for different types of planets}\label{sec::escape}

To estimate the atmospheric mass loss rates for specific planets, I employ the grid of hydrodynamic upper atmosphere models presented in \cite{kubyshkina2018grid} and \cite{kubyshkina2021RN}. It is based on the 1D model describing pure hydrogen atmosphere heated by stellar XUV. The latter is considered as X-ray and EUV fluxes \citep[as in][]{SF2011} integrated over the spectrum and emitted at the single wavelengths of 5 and 62~nm. The lower boundary is the photosphere of a planet, where the atmosphere is assumed to consist of molecular hydrogen, the outflow velocity is set to zero, temperature equals the equilibrium temperature of the planet (with zero albedos), and the atmospheric pressure is set according to lower atmosphere models \citep{cubillos2017}. The upper boundary is the planetary Roche lobe where the flow is assumed continuous. The code solves the mass, momentum, and energy conservation equations together with continuity equations accounting for dissociation, recombination, ionization, and collisional, Ly$-\alpha$, and $H_3^+$-cooling.

This model was used to perform upper atmosphere modeling for the large grid of planets orbiting stars of different masses and XUV luminosities. The grid covers planetary masses ($M_{\rm pl}$) of 1-109~$M_{\oplus}$, radii ($R_{\rm pl}$) of 1-10~$R_{\oplus}$ and orbital separations ($d_0$) corresponding to equilibrium temperatures ($T_{\rm eq}$) of 300-2000~K. The relation between $d_0$ and $T_{\rm eq}$ for a star of specific mass is set up as an average value based on stellar isochrones \citep{choi2016}. Masses of host stars ($M_*$) range between 0.4-1.3~$M_{\odot}$, and for each star, we consider not less than 3 different XUV luminosities between $\sim$0.5 and $10^4$ of the present Sun, scaled for particular $M_*$. More detail on the model and the grid structure can be found in \cite{kubyshkina2018grid}.

\begin{figure*}[t]
\centerline{\includegraphics[width=0.6\hsize]{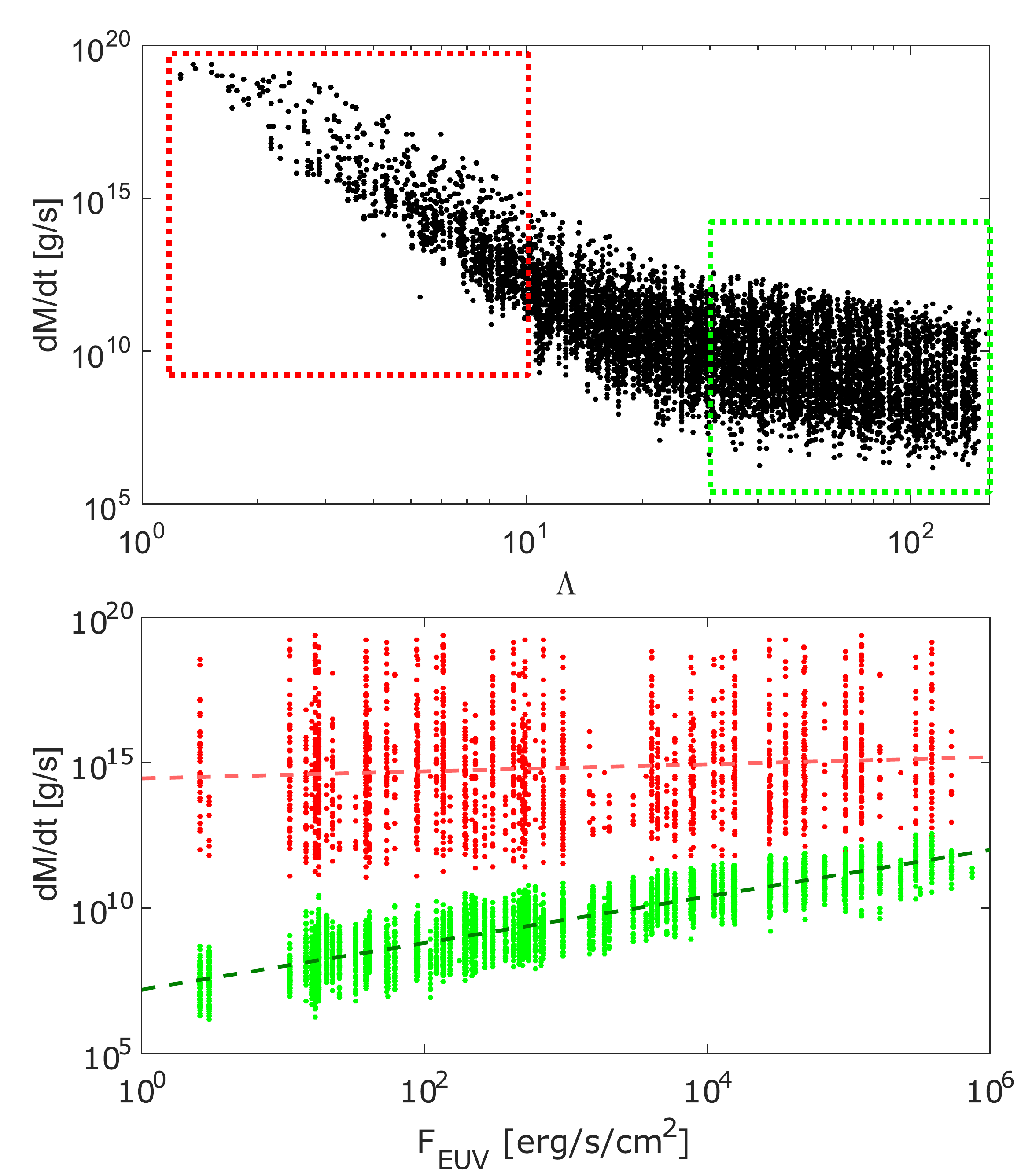}}
\caption{Atmospheric mass loss rates predicted by the hydrodynamic model against the generalized Jeans escape parameter (top panel) and the EUV flux exposed at the planet (bottom panel). Each point represents the specific planet from the grid. The red (low $\Lambda$) and the green (high $\Lambda$) rectangles in the top panel highlight the groups of planets shown in red and green colors in the bottom panel, respectively. The coral and the forest green dashed lines in the bottom panel show the linear fits to these groups of points in log-log space.\label{fig::mdot}}
\end{figure*}

In Figure~\ref{fig::mdot} (top panel), I present the atmospheric mass loss rates for all planets in the grid ($\sim$10000) against the generalized Jeans escape parameter \citep{fossati2017}

\begin{equation}
    \Lambda = \frac{G M_{\rm pl}m_{\rm H}}{k_{\rm b}T_{\rm eq}R_{\rm pl}},
\end{equation}

\noindent where $G$, $m_{\rm H}$, $k_{\rm b}$ are the gravitational constant, hydrogen mass, and the Boltzmann constant. This parameter represents the relation of the gravitational energy of a planet to its thermal energy. Thus, points on the left-hand side of the diagram represent in general low-density and/or very hot planets, while points on the right-hand side correspond to heavier, cooler, and more compact planets. These two groups form two distinct brunches of $\frac{dM}{dt}(\Lambda)$ dependence. The mass loss decreases steeply with increasing gravity and decreasing thermal energy of the planet for low-$\Lambda$ planets, and the slope becomes more moderate for high-$\Lambda$ planets. I highlight the typical representatives of these two groups with the red and the green rectangles, omitting the transition region with $\Lambda$ of 10-30.

In the bottom panel of Figure~\ref{fig::mdot}, I examine how the atmospheric mass loss from planets in these two groups depends on stellar EUV flux at planetary orbit ($F_{\rm EUV}$). I have chosen the EUV flux instead of the full XUV, as for most of the planets in the grid the effect of the EUV part of the spectrum dominates the one of X-ray, and the dependencies look clearer against $F_{\rm EUV}$. In addition, the X-ray to EUV relation by \citet{SF2011} used in the grid was recently shown to overestimate the EUV emission for stars with high X-ray fluxes \citep[see, e.g,][]{chadney2015,johnstone2021}. Thus, for a given $F_{\rm EUV}$ the total XUV fluxes can vary for different approximations, and $F_{\rm EUV}$ provides a better reference for comparisons. Each model in Figure~\ref{fig::mdot}, however, includes both EUV and X-ray fluxes.
For clarity, I add to the plot the linear approximations in log-log space. One can see that planets in the low-$\Lambda$ group (red points) show weak dependence on $F_{\rm EUV}$, with average mass loss rate changing a few times while EUV flux changes by $\sim$6 orders of magnitude. This can be interpreted in a way, that atmospheric escape from such planets is mainly driven by the own thermal energy of a planet rather than stellar XUV, which provides only a small addition to the atmospheric mass loss rate. The mass loss rates in this group are comparable (order of magnitude) to those predicted by the core-powered mass loss model \citep{gupta_schlichting2019} for the same lower boundary conditions \citep{kubyshkina2021RN}.

On a contrary, atmospheric mass loss rates from high-$\Lambda$ planets depend on $F_{\rm EUV}$ nearly linearly in log-log space (i.e., $\frac{dM}{dt}\sim F_{\rm EUV}^{\alpha}$, where $\alpha$ is close to 1). For comparison, the energy-limited approximation predicts the linear dependence. This holds for hydrodynamic models at high $\Lambda$ as well, but for fluxes higher than $\sim$100~${\rm erg/s/cm^2}$. For lower EUV fluxes, $\frac{dM}{dt}$ decreases slower with decreasing $F_{\rm EUV}$ than linear law. Thus, the atmospheric mass loss rates predicted for this group by the hydrodynamic model and the energy-limited approximation are at a similar level except for planets under low insolation \citep{kubyshkina2021RN,krenn2021}.

To summarize, according to the hydrodynamic model the atmospheric mass loss rate does not depend on stellar XUV the same way for all planets. Instead, the level of XUV (hence, any variations in it) is of crucial importance only for high-$\Lambda$ planets (relatively massive and compact).

\section{Atmospheric escape in context of planetary evolution}\label{sec::escape_evol}

Low to intermediate-mass planets can change their parameters (in particular, radii) significantly throughout their lives due to the atmospheric mass loss and the thermal evolution of a planet (i.e., relaxation of post-formation luminosity). The former is most relevant for planets at close-in orbits, while the latter occurs at all planets. 

\begin{figure*}[h]
\centerline{\includegraphics[width=0.6\hsize]{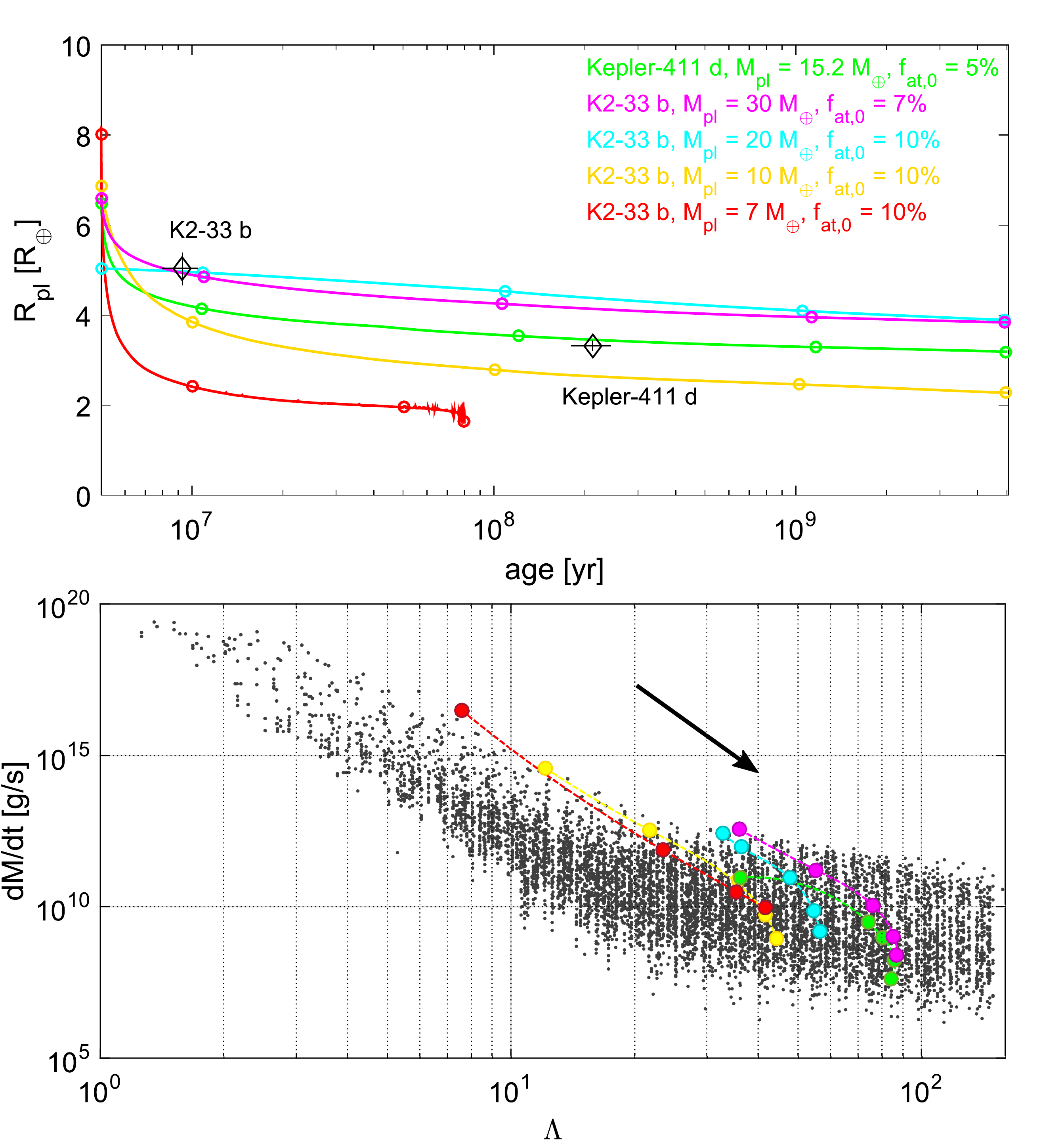}}
\caption{Top panel: evolutionary tracks of planetary radii against time for Kepler-411~d (green) and a few models reproducing K2-33~b assuming different planetary masses of 30 (magenta), 20 (cyan), 10 (yellow), and 7~$M_{\oplus}$ (red). The circles denote points in the tracks closest to 5, 10, and 100~Myr and 1 and 5~Gyr, except for the case of 7~$M_{\oplus}$ version of K2-33~b, where the planet loses its atmosphere before 100~Myr. In this case, the circles correspond to 5, 10, 50, and 80~Myr. The present-day parameters of K2-33~b and Kepler-411~d are shown by black diamonds, and the black lines denote the corresponding uncertainties. Bottom panel: the atmospheric mass loss rates against the generalized Jeans escape parameter with overplotted ($\frac{dM}{dt}(\Lambda)$) time snapshots of evolving planets from the top panel. The color code is the same as in the top panel, and each circle corresponds to the one on the $R_{\rm pl}(age)$ track, with the earliest time on the left. The lines in the bottom panel do not represent the actual $\frac{dM}{dt}(age)$ track. \label{fig::mdot_evol}}
\end{figure*}

In Figure~\ref{fig::mdot_evol}, I present the evolution of a few particular planets (previously considered in \citealt{kubyshkina2021fat}) and how their position in the $\frac{dM}{dt}(\Lambda)$ diagram changes with time. To build the planetary evolution tracks, I use the framework presented in \citet{kubyshkina2020fat}. It combines the thermal evolution of the planet performed with Modules for Experiments in Stellar Astrophysics
\citep[MESA][]{Paxton2011, Paxton2013, Paxton2015, Paxton2018} with the atmospheric mass loss prescription based on hydrodynamic models presented in Section~\ref{sec::escape} \citep{kubyshkina2018app}. To model the evolution of stellar XUV luminosity, I employ the Mors code \citep{johnstone2021} and fit the parameters of host stars at the given age of the systems.

First, I consider the evolution of Kepler-411~d, the $15.2\pm5.1$~$M_{\oplus}$ planet orbiting $0.870\pm0.039$~$M_{\odot}$ star at $\sim0.28$~AU \citep{sun2019}. I start the evolution at the time of protoplanetary disk dispersal (which is set at 5~Myr and has a minor effect on results), assuming the initial atmospheric mass fraction of 5\%. It allows reproducing parameters of the planet at the given age of the system \citep{kubyshkina2021fat}. Despite the host star being young ($212\pm31$~Myr) and active ($L_{\rm XUV}\sim 9.35\times 10^{28}$~erg/s, which corresponds to $F_{\rm XUV}\sim 427$~erg/s/cm$^2$ at the planetary orbit; see \citealt{xu2021,johnstone2021}), the planet does not experience an atmospheric mass loss strong enough to alter its atmospheric mass substantially throughout the lifetime. Overall, it loses less than 1\% of its atmosphere, and its evolution (in particular, $R_{\rm pl}(age)$, green line in the top panel of Figure~\ref{fig::mdot_evol}) is mainly controlled by thermal evolution. In all cases considered here, the initial entropy is taken of 8.5~$k_{\rm b}/baryon$, unless explicitly specified otherwise. All the effects from different initial entropies disappear within a few tens of Myr. 

Next, I consider a few model planets in a position of K2-33~b, a very young planet ($9.3_{-1.3}^{+1.1}$~Myr) orbiting close-in to its host star. The orbital separation of $\sim0.0409$~AU and the stellar parameters ($M_*\sim$0.56~$M_{\odot}$, $R_* = \sim1.05R_{\odot}$, $T_{\rm eff} = 3540$~K, \citealt{mann2016}) result in $T_{\rm eq}\sim850$~K, and XUV flux at the planetary orbit of $\sim6.4\times10^4$~erg/s/cm$^2$. This suggests that the planet can undergo a strong atmospheric mass loss, with actual value depending on $M_{\rm pl}$. For K2-33~b, only a conservative upper limit on planetary mass is available, $M_{\rm pl}\leq 3.7M_{\rm jup}$. However, the size of the planet ($R_{pl} = 5.04^{+0.34}_{-0.37}$~$R_{\oplus}$) suggests that it is unlikely heavier than Saturn, i.e., $\sim95$~$M_{\oplus}$. Here, I show the tracks for planets of 30 (magenta line in the top panel of Figure~\ref{fig::mdot_evol}), 20 (cyan), 10 (yellow), and 7~$M_{\oplus}$ (red), starting their evolution with initial atmospheric mass fractions of 7\% (for the heaviest) or 10\%. Each corresponds to a different level of atmospheric mass loss. In particular, our models predict that 30~$M_{\oplus}$ planets loses $\sim2\%$, the 20~$M_{\oplus}$ planet loses $\sim13\%$, and the 10~$M_{\oplus}$ planet loses $\sim93\%$ of its initial atmosphere. The two former tracks reproduce the radius of K2-33~b at the given age of the system within the observational uncertainties, and the 10~$M_{\oplus}$ planet could only reproduce it if assuming a very low initial luminosity (below 7~$k_{\rm b}/baryon$).
The last model planet, with a mass of 7~$M_{\oplus}$, loses its atmosphere within 100~Myr and does not reproduce the radius of K2-33~b at its age for any initial conditions. It was included to demonstrate more of different possible scenarios. For the same purpose, for 20~$M_{\oplus}$ model I chose to show the case with low initial entropy of 7.5~$k_{\rm b}/baryon$.

In the bottom panel of Figure~\ref{fig::mdot_evol}, I plot the atmospheric mass loss rates at the ages marked by circles in the planetary evolutionary tracks in the top panel (colors for each planet are preserved) over the $\frac{dM}{dt}(\Lambda)$ diagram from Figure~\ref{fig::mdot}.
For most planets considered here, the largest change in $\frac{dM}{dt}$ (as the largest change in $R_{\rm pl}$) occurs within the first 10-100~Myr of their lifetime, including the case of Kepler-411~d, where the atmospheric mass loss is minor. The only exclusion is the 20~$M_{\oplus}$ planet, which starts its evolution with the low initial temperature of the rocky core.
As $R_{\rm pl}$ decrease due to the cooling of planetary core and atmospheric mass loss, all planets move within the diagram from lower to higher $\Lambda$ and towards smaller atmospheric mass loss rates. Changes in $\frac{dM}{dt}$ remain significant also after the age of 100~Myr, when the radii of planets (hence, $\Lambda$) change only slightly, due to the decreasing stellar XUV.

Though these patterns are the same for all planets considered here, the starting point depends on the mass and temperature (both core temperature and $T_{\rm eq}$) of the planet. Thus, low-mass and/or hot planets start their evolution on the left-hand side of the diagram, where the atmospheric mass loss is mainly driven by high thermal energy and low gravity of the planet and weakly depends on stellar XUV. However, they move to the high-$\Lambda$ regime within the first 100~Myr. On the other hand, more massive and/or cooler planets start their evolution already on the right-hand side of the diagram or in the transition region, where the atmospheric mass loss rates depend strongly on XUV. This, however, does not automatically mean that stellar XUV has a strong impact on the evolution of planetary radius: thus, in the case of Kepler-411~d, the atmospheric mass loss depends strongly on XUV but appears to be too low to remove a significant part of the atmosphere and the $R_{\rm pl}(age)$ is mainly controlled by thermal evolution.

\section{Stellar rotation and activity}\label{sec::stellar_rot}

It is known, that at the beginning of evolution, stars remain in the saturation regime, where $L_{\rm XUV}$ depends weakly on stellar rotation. As rotation slows down and reaches some critical value, the star leaves this regime and stellar $\frac{L_{\rm X}}{L_{\rm bol}}$ decays as $\sim age^{-\alpha}$, where $\alpha$ is a constant ranging between $\sim1.1-2.0$ \citep[as was resolved in a range of studies][]{wright2011,jackson2012,magaudda2020}. However, different stars are not born with the same rotation rate, and those that rotate faster remain longer in the saturation regime compared to slow rotators. This leads to the wide spread in stellar $L_{\rm XUV}$ at early ages before 1-2~Gyr. This spread looks differently for stars of different masses: lower mass stars evolve slower than heavier, so the spread, in this case, occurs later and is less pronounced \citep[see, e.g.,][]{matt2015,johnstone2021}.

\begin{figure}[t]
\centerline{\includegraphics[width=\hsize]{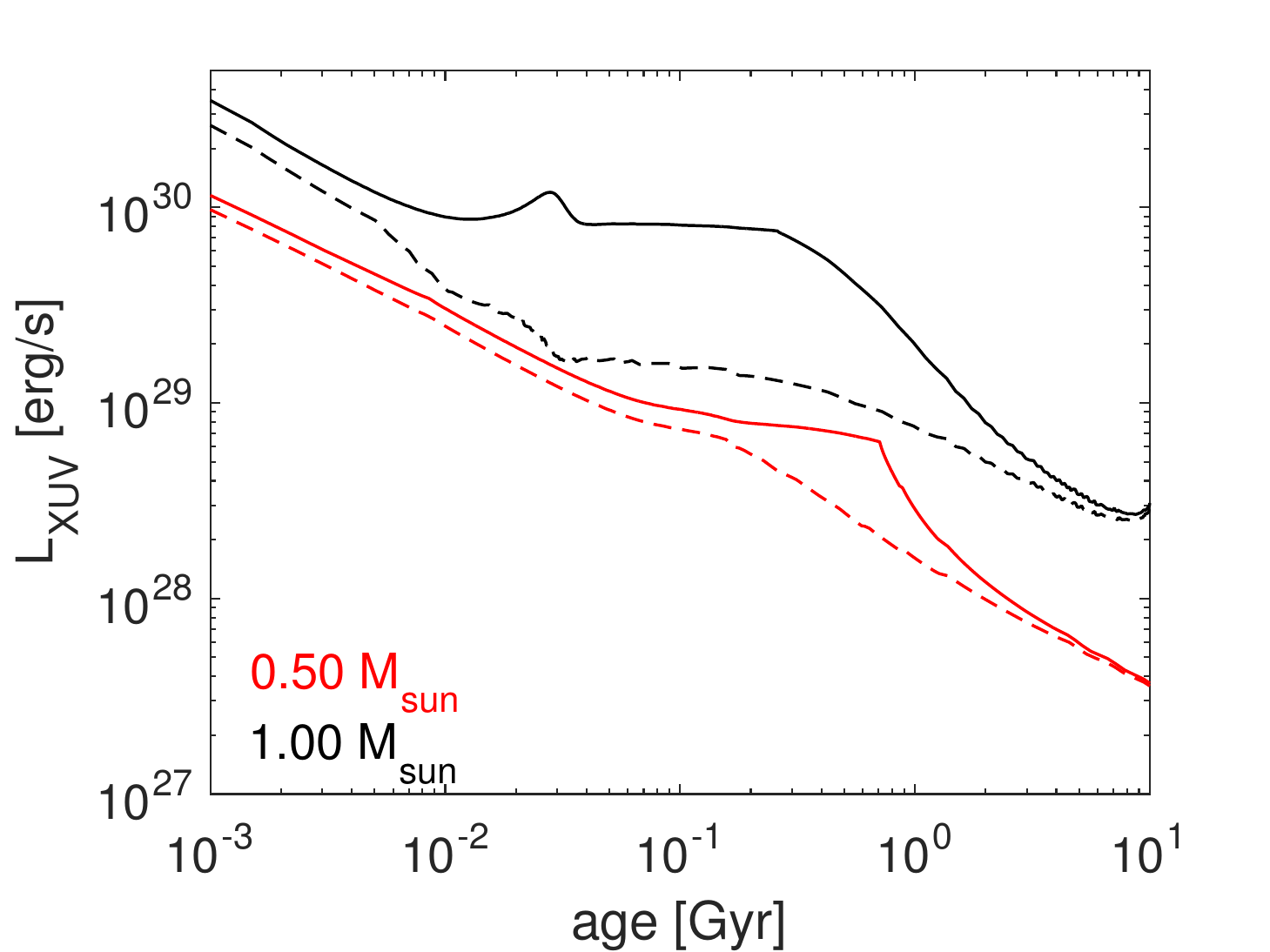}}
\caption{$L_{\rm XUV}$ against time for 0.5~$M_{\odot}$ (red) and 1.0~$M_{\odot}$ (black) stars evolving as fast ($P_{\rm rot,150} = 0.5/1$~days, solid lines) or slow ($P_{\rm rot,150} = 7$~days, dashed lines) rotators \citep{johnstone2021}. \label{fig::xuv}}
\end{figure}

In Figure~\ref{fig::xuv}, I show four stellar $L_{\rm XUV}(age)$ models produced using the Mors code \citep{johnstone2021}, which is based on fitting parameters of stellar rotation and XUV models on large stellar samples covering ages starting from 2~Myr and stellar masses between 0.1 and 1.25~$M_{\odot}$. Here, I consider stars of 0.5 (red lines in Figure~\ref{fig::xuv}) and 1.0~$M_{\odot}$ (black lines) evolving as fast (solid lines) or slow rotators (dashed lines). As a proxy for the rotation type, I use the rotation period at 150~Myr: for the slow rotators I assume $P_{\rm rot,150} = 7$~days, while for the fast rotators I adopt $P_{\rm rot,150}$ of 0.5 and 1 day for 0.5~$M_{\odot}$ and 1.0~$M_{\odot}$ stars, respectively. These values represent well the slow and fast edges of the rotation distributions for stars of similar ages and masses according to observations \citep[see, e.g.,][]{johnstone2015}.
For 0.5~$M_{\odot}$ and 1.0~$M_{\odot}$ stars, the difference in $L_{\rm XUV}$ between the fast and the slow rotating stars is much more pronounced for the heavier one. At the same time, $L_{\rm XUV}$ is always higher for solar mass star compared to 0.5~$M_{\odot}$ star, also if considering the slow rotating 1.0~$M_{\odot}$ star against the fast rotating 0.5~$M_{\odot}$ star. One should note, however, that lower-mass stars are typically cooler than heavier ones, and their habitable zones (i.e., $T_{\rm eq}\sim$300~K) are located much closer to the star. This leads to that at the orbits corresponding to the same $T_{\rm eq}$, planets receive more XUV around lower mass stars \citep{johnstone2021}.

\section{Relevance for planetary population}\label{sec::population}

\begin{figure*}[h]
\centerline{\includegraphics[width=0.5\hsize]{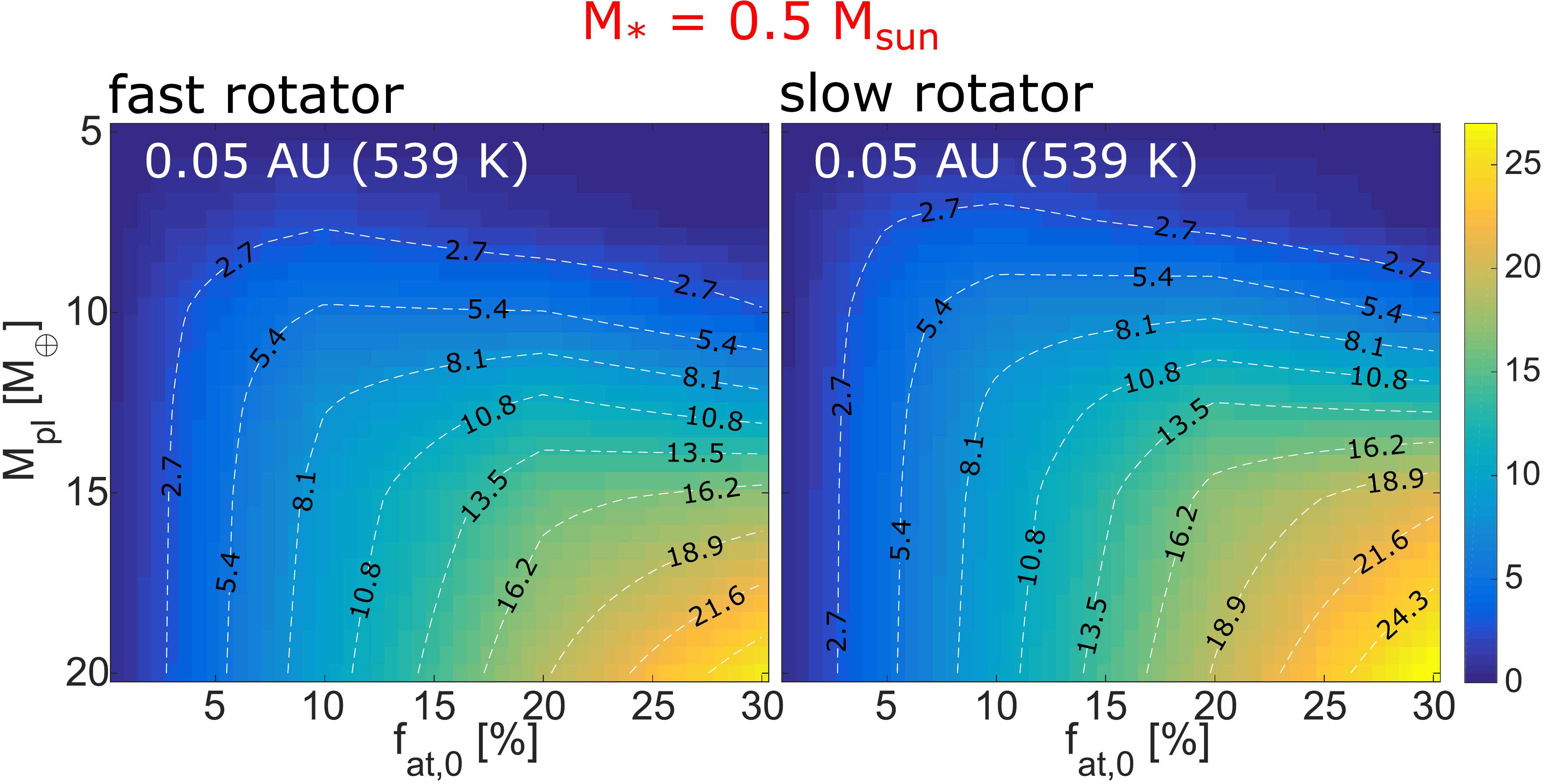} \includegraphics[width=0.5\hsize]{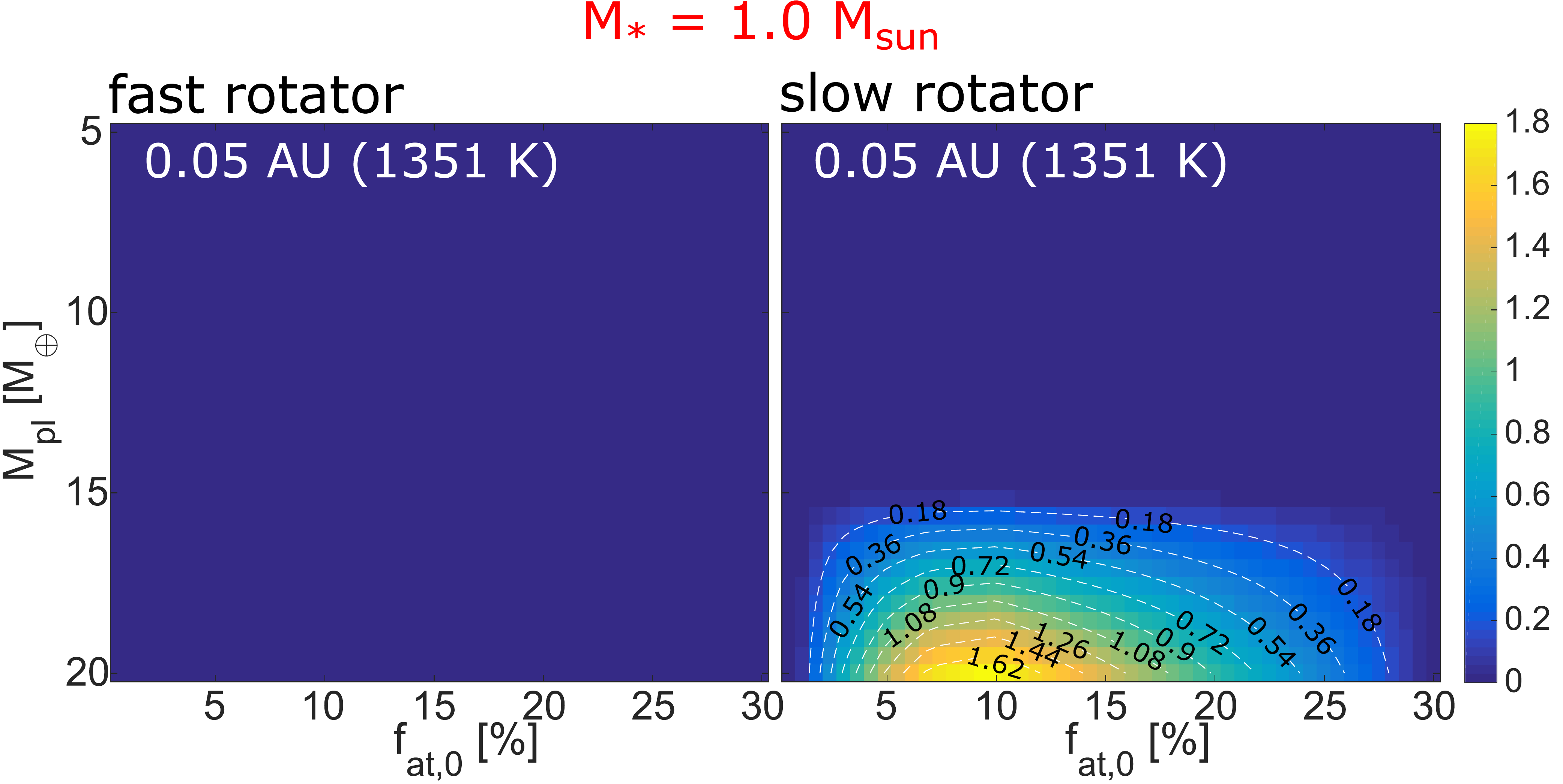}}
\caption{The atmospheric mass fraction after 5~Gyr evolution against planetary mass (Y-axis) and initial atmospheric mass fraction (X-axis) at 0.05~AU. The left two panels correspond to 0.5~$M_{\odot}$ star, and the right two columns -- to 1.0~~$M_{\odot}$ star. In each pair of panels, the left one shows the final $f_{\rm at}$ distribution assuming that the star has evolved as a fast rotator, and the right one assuming a slow rotator. The corresponding equilibrium temperatures for each stellar mass are given in the plot. \label{fig::fat_dist}}
\end{figure*}

Employing the stellar evolution models from Section~\ref{sec::stellar_rot}, I compare here evolution of planets around stars of different masses and rotation histories. For that, I run the atmospheric evolution (as described in Section~\ref{sec::escape_evol}) for planets with masses between 5-20~$M_{\oplus}$, considering for each mass the initial atmospheric mass fractions between 0.5 and 30\% and orbital separations between 0.03-0.5~AU. Lower mass planets in close-in orbits (where the effect from various stellar histories is expected to be relevant) tend to lose their primordial atmospheres well before the age of 1~Gyr and are therefore excluded from consideration. In Figure~\ref{fig::fat_dist}, I show the final atmospheric mass fractions that planets preserve after 5~Gyr evolution against planetary mass and initial atmospheric mass fraction at the orbital separation of 0.05~AU for both stellar masses and both rotation scenarios.

Planets evolving at the same orbital separation around 1.0~$M_{\odot}$ star preserve much less of the atmosphere compared to 0.5~$M_{\odot}$ star, due to higher $F_{\rm XUV}$ and $T_{\rm eq}$, and the difference between planets evolved at close-in orbits around the fast and the slow rotators is larger for the heavier star.
With increasing $d_0$ (decreasing $T_{\rm eq}$), the effect from atmospheric mass loss decreases, and the atmospheric mass fractions for planets heavier than a certain value (different for different $d_0$) remain nearly unchanged, independent from the stellar rotation history. Therefore, the difference between planetary populations evolved at the slow and the fast rotator becomes insignificant. For stars of different masses, the border value of $d_0$ is different, but the border temperatures are similar -- about 500~K.

\begin{figure*}[t]
\centerline{\includegraphics[width=0.5\hsize]{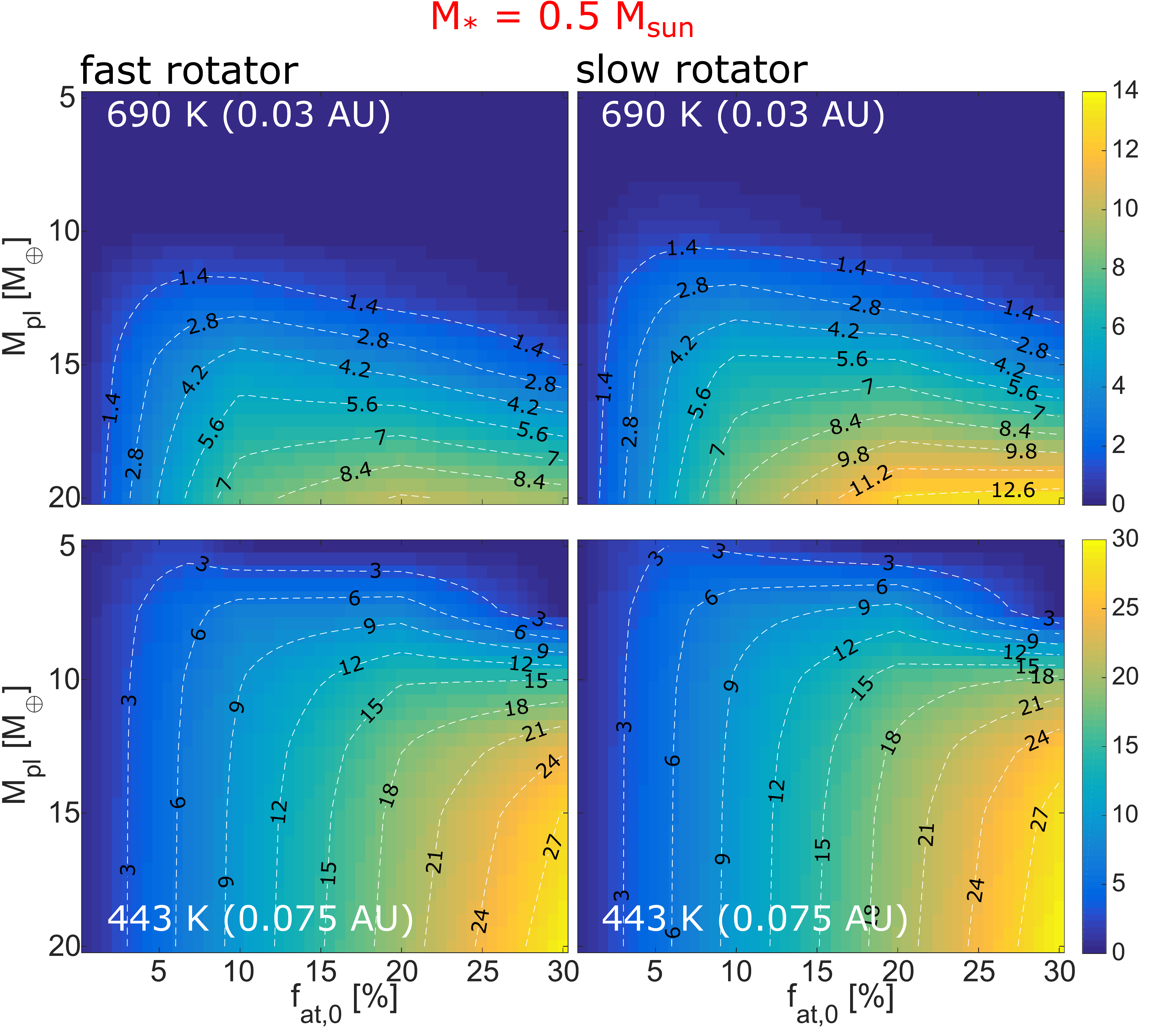} \includegraphics[width=0.5\hsize]{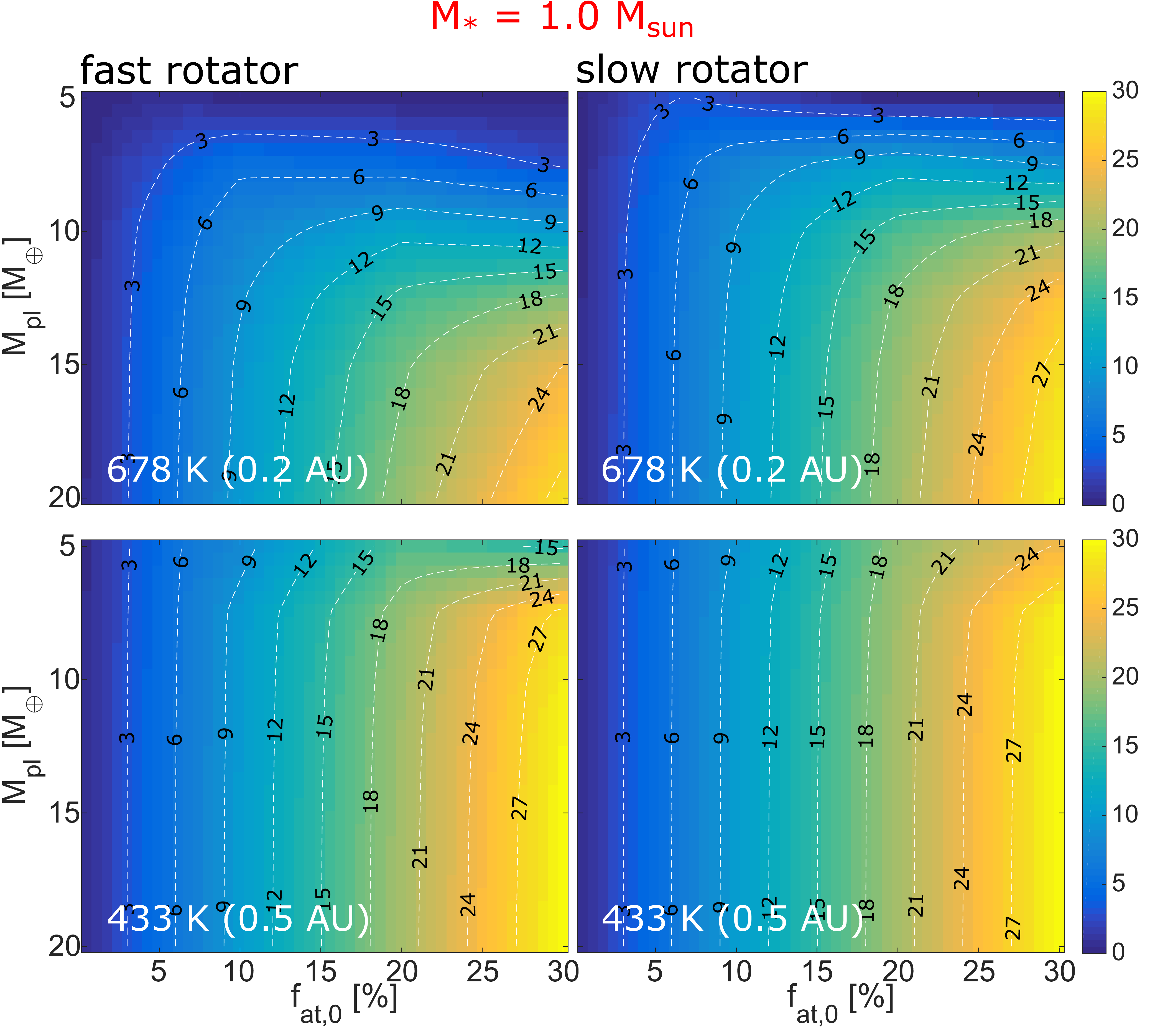}}
\caption{Same as in Figure~\ref{fig::fat_dist}, but for 0.03 and 0.075~AU for $M_* = 0.5 M_{\odot}$ (left), and for 0.2 and 0.5~AU for $M_* = 1.0 M_{\odot}$ (right). The orbital separations in the same row correspond to the similar averaged equilibrium temperatures at each star. \label{fig::fat_teq}}
\end{figure*}

In Figure~\ref{fig::fat_teq}, I compare the same distributions as in Figure~\ref{fig::fat_dist} but for orbital separations corresponding to similar $T_{\rm eq}$ around stars of different masses. I pick the two $T_{eq}$ values of $\sim$680~K (top row) and $\sim$430~K (bottom row). In this case, planets with similar $T_{\rm eq}$ receive more $F_{\rm XUV}$ around the lower mass star, due to the closer orbit. At $\sim$100~Myr, the difference is about 5.1 times, if considering the fast rotators. This leads to substantially smaller final atmospheres at the given $T_{\rm eq}$ in case of 0.5~$M_{\odot}$ star compared to 1.0~$M_{\odot}$ star. 
For similar $T_{\rm eq}$, the distributions of $f_{\rm at,5Gyr}$ around stars of different masses show similar differences between cases of fast and slow rotators, despite the actual difference between $L_{\rm XUV}(age)$ tracks is smaller for 0.5~$M_{\odot}$ star. This can be partly due to higher $F_{\rm XUV}$ at the given $T_{\rm eq}$, and partly because for low-mass planets $\frac{dM}{dt}$ during the first $\sim$100~Myr is controlled mainly by the own thermal energy of the planet and weakly depends on XUV, as discussed in Section~\ref{sec::escape_evol}. The latter could diminish the importance of the difference between slow and fast rotators through this period, which is particularly large for the solar-mass star (see Figure~\ref{fig::xuv}).
For the whole population, however, the difference between fast and slow rotating stars remains more important for the heavier star, as for 0.5~$M_{\odot}$ star, all the relevant orbits are within 0.05~AU, while for the solar mass star the effect presents as far as 0.2~AU.

At specific orbit, the difference between $f_{\rm at,5Gyr}$ of planets orbiting fast and slow rotators depends on planetary parameters. For heavier planets starting their evolution with compact envelopes (see 15-20~$M_{\oplus}$ and $f_{\rm at,0}\leq\sim7\%$ in Figure~\ref{fig::fat_teq}), the difference is minor or absent, but it increases with decreasing $M_{\rm pl}$ and increasing $f_{\rm at,0}$. So, the effect from various stellar rotation histories should be most relevant for low-mass planets with substantial $f_{\rm at,0}$. However, there are two factors that can affect this conclusion. First, low-mass planets ($M_{\rm pl}\leq\sim10$~$M_{\oplus}$) at close-in orbits tend to lose their atmospheres within the first Gyr of evolution independent of the host star rotation type. Therefore, after $\sim1~Gyr$, one can not see any difference between fast and slow rotating stars. Second, according to formation models, close-in low-mass planets are unlikely to accrete a substantial primordial atmosphere \citep[see, e.g.,][and references therein]{mordasini2020}. Therefore, the most effect from uncertain stellar XUV evolution is likely to be seen at intermediate-mass planets starting their evolution with $f_{\rm at,0}\geq10\%$.

\section{Conclusions}\label{sec::conclusions}

The wide spread possible stellar XUV luminosities at early ages represent an uncertain factor in studies of the observed population of exoplanets and the ways it could be formed. Here, I estimate how large an effect it may have and for which stellar and planetary parameters it is most relevant.

The variability in possible $L_{\rm XUV}(age)$ is important for planetary population (radii distribution) in case if the latter was mainly formed by atmospheric mass loss driven by stellar irradiation.
Hydrodynamic modeling in Section~\ref{sec::escape} suggests that the main driving mechanism of the atmospheric escape can be different for different planets and also for a specific planet at different stages in evolution. The atmospheric mass loss depends strongly on stellar XUV for planets with $\Lambda$ above 20-30, which implies relatively heavy, compact, and/or cool planets. For planets heavier than 20-40~$M_{\oplus}$ it is typically the case for most of their lives, while for the lighter planets $\Lambda$ can remain below 20 for the first tens/hundreds Myr. At this time they undergo an intense atmospheric loss driven by their own thermal energy, which can be decisive for their final position in the radius-period diagram. After that, however, they switch to the high-$\Lambda$ regime with mass loss dependent on XUV. 

The evolutionary modeling performed for wide range of planets reveals that the effect from different stellar rotation histories is most relevant for planets with $T_{\rm eq}$ above 500~K, with intermediate masses and large primordial atmospheres. Despite the difference in populations evolved at fast and slow rotating stars is similar for the specific $T_{\rm eq}$ around stars of different masses, the effect will likely be more important for heavier stars, because of the larger temperatures at close-in orbits.



\section*{Acknowledgments}
This project has received funding from the \fundingAgency{European Research Council} (ERC) under the European Union’s Horizon 2020 research and innovation programme (grant agreement \fundingNumber{No 817540}, ASTROFLOW).

\subsection*{Financial disclosure}

None reported.

\subsection*{Conflict of interest}

The authors declare no potential conflict of interests.

\bibliography{stellar_rot}%

\end{document}